\documentclass[useAMS,usenatbib,a4]{mn2e}

\usepackage{float}
\usepackage{graphicx}
\usepackage{amssymb}
\usepackage{amsmath}
\usepackage{datetime}
\usepackage[normalem]{ulem}
\usepackage{times}
\usepackage[export]{adjustbox}

\usepackage{color}
\usepackage{soul}
\definecolor{stan}{rgb}{0,0,1}

\definecolor{jon}{rgb}{0,1,0}

\definecolor{nir}{rgb}{1,0,0}

\def\<<{{\ll}}
\def\>>{{\gg}}

\def\spose#1{\hbox to 0pt{#1\hss}}
\def\ltwig{\mathrel{\spose{\lower 3pt\hbox{$\mathchar"218$}}
     R_{\rm A}ise 2.0pt\hbox{$\mathchar"13C$}}}
\def\gtwig{\mathrel{\spose{\lower 3pt\hbox{$\mathchar"218$}}
     R_{\rm A}ise 2.0pt\hbox{$\mathchar"13E$}}}
\def\+/-{{\pm}}
\def\=={{\equiv}}

\def\Rw{R_{w}}
\def\Rstar{R_{\ast}}

\def\Teff{T_{\rm eff}}

\def\Mdot{\dot M}

\def\solar{\odot}
\def\Msun{M_{\solar}}

\def\Rsun{R_{\solar}}
\def\Lsun{L_{\solar}}
\def\Tsun{T_{\solar}}

\newcommand{\beq}{\begin{equation}}
\newcommand{\eeq}{\end{equation}}
\newcommand{\beqa}{\begin{eqnarray}}
\newcommand{\eeqa}{\end{eqnarray}}

\begin{document}

\title[Spectral Temperature of Giant Eruptions]
{
The Spectral Temperature of Optically Thick Outflows with Application to
Light Echo Spectra from $\eta$~Carinae's Giant Eruption
}

 \author[Owocki and Shaviv]
{
\vbox{
Stanley P.\ Owocki$^1$\thanks{Email: owocki@udel.edu}
and Nir J. Shaviv$^2$\thanks{Email: shaviv@phys.huji.ac.il}
}
\\ $^1$ Department of Physics and Astronomy, Bartol Research Institute,
 University of Delaware, Newark, DE 19716, USA
 \\ $^2$ Racah Institute of Physics,
Hebrew University,  Giv'at Ram Jerusalem 91904 Israel
}

\maketitle

\begin{abstract}
 The detection by \citet{Rest12} of light echoes from $\eta$\,Carinae has provided important new observational constraints on the nature of its 1840's era giant eruption. Spectra of the echoes suggest a relatively cool spectral temperature of about 5500K, lower than the lower limit of about 7000K suggested in the optically thick wind outflow analysis of \citet{Davidson87}.  This has lead to a debate about the viability of this steady wind model relative to alternative, explosive scenarios. Here we present an updated analysis of the wind outflow model using newer low-temperature opacity tabulations and accounting for the stronger mass loss implied by the $>$10 Msun mass now inferred for the Homunculus.  A major conclusion is that, because of the sharp drop in opacity due to free electron recombination for $T<$6500K, a low temperature of about 5000K is compatible with, and indeed expected from, a wind with the extreme mass loss inferred for the eruption. Within a spherical gray model in radiative equilibrium, we derive spectral energy distributions for various assumptions for the opacity variation of the wind, providing a basis for comparisons with observed light echo spectra. The scaling results here are also potentially relevant for other highly optically thick outflows, including those from classical novae, giant eruptions of LBVs and SN Type IIn precursors. A broader issue therefore remains whether the complex, variable features observed from such eruptions are better understood in terms of a steady or explosive paradigm, or perhaps a balance of these idealizations.
 \end{abstract}


\section{Introduction}

The high luminosities of massive stars drive powerful stellar winds, 
with speeds of several thousand km/s, and mass loss rates a billion times that of the solar wind
\citep[see, e.g., the review by][and references therein]{Puls08}.
For stars of spectral type O and B, the winds generally remain optically thin in the continuum, 
with the observed spectral energy distribution (SED) thus characterized by the high temperature (20,000-50,000\,K) of the underlying star.

The much stronger winds of  Wolf-Rayet (WR) stars remain optically thick well into the supersonic outflow, with an effective ``wind photospheric radius'' $\Rw$ that is typically several times the hydrostatic stellar core radius $\Rstar$.
This has a profound effect on the emergent spectra, which become characterized by wind-broadened emission lines of HeII and moderately ionized stages of N, C or O, representing the so-called WN, WC or WO spectral types.
Nonetheless, because of the large luminosities, combined with relatively small core radii, the overall SED of such WR stars still suggest quite high  spectral temperatures, again several times 10,000\,K.

The so-called S-Doradus type Luminous Blue Variable (LBV) stars  exhibit horizontal excursions to cooler temperatures at roughly constant bolometric luminosity, and it was initially thought these might arise from an increased size of a wind photosphere associated with enhanced mass loss 
\citep{Humphreys94}.
Subsequent analyses \citep{deKoter96} have shown, however, the inferred mass loss rates ($< 10^{-5} \Msun$/yr) are well below the $\sim 10^{-3} \Msun$/yr needed (see equation \ref{eq:Tke}) for a wind photosphere with effective temperature below $\sim 10,000$\,K.

However,  the ``giant eruption'' class of LBV stars can indeed have mass loss rates that meet or exceed such large values, suggesting then relatively cool effective temperatures below $10,000$\,K, which are indeed  observed in the ``supernova imposters'' that are now thought to be examples of giant eruption LBVs in external galaxies \citep{Smith11}.
%
But an early theoretical analysis by \citet[][hereafter D87, see \S \ref{sec:D87anal}]{Davidson87} concluded that, due to a sharp drop in opacity associated with the recombination of free electrons, the spectral temperature would reach an apparent minimum around 7000\,K, even for cases with  strong mass loss.

In recent years the detection by \citet{Rest12} of light echoes from the 1840's giant eruption of $\eta$\,Carinae has provided important new observational constraints on the nature of such giant eruptions. In particular, spectra of these echoes suggest a relative cool spectral temperature of $\sim$5000-6000\,K. Noting that this is lower than the $\sim$7000\,K lower-limit cited by the optically thick wind outflow analysis of
 D87,
 \citet{Rest12} have suggested this argues against such a steady wind scenario, and instead may favor an {\em explosive} model for the giant eruption in $\eta$\,Carinae, and by extension even other LBV's.
 A similar debate exists for interpreting the nature of the eruptive precursors that supernovae of Type IIn have months before their final event \citep{Ofek14, Smith14}, and also in classical novae eruptions \citep{Friedjung11}. 
 
 In this paper we focus on the specific question of whether a sufficiently strong steady wind outflow could, in principle, produce the low temperatures inferred from these light  echoes.
Even within this steady-state context, a full model for the emergent spectrum of an optically thick wind outflow should treat the wind ionization balance and its associated effect on the opacity. In principle this should include non-LTE treatment of the scattering and absorption from both line and continuum in an expanding wind outflow, using for example  stellar wind NLTE transport codes like CMFGEN \citep{Hillier12, Hillier99}, FASTWIND \citep{Puls05}, or WM-Basic \citep{Pauldrach01}.
However, these codes were developed to model massive-star winds for relatively high temperatures, typically above about 10,000\,K, and so in their standard form are not readily suited for treatment of the cooler temperatures that can develop in the outer regions of very massive outflows, such as occur during the giant eruptions of LBV stars.

As an initial alternative, we explore here much simpler analyses aimed at determining semi-analytically the scaling of spectral temperature with mass loss. In section \ref{sec:BBwind} we first use an idealized LTE blackbody (BB) wind model with fixed wind opacity to obtain a simple, explicit relation (equation \ref{eq:Tke}) for effective temperature $T_{\rm eff}$ in terms of opacity $\kappa$, stellar luminosity $L$, and the ratio of wind mass-loss-rate to flow speed, $\Mdot/V$.
In section \ref{sec:sgre} we extend this to account for sphericity of the wind, within a spherical, gray, radiative equilibrium model with a photon mean-free-path that varies as a power-law in radius, $\ell \sim 1/\kappa \rho \sim r^{n}$.
For small $n =2$, the emergent spectrum (see figure \ref{fig:fx}) peaks at lower energy, with a non-BB high-energy tail; but in the limit of large $n$ one recovers the BB results of section \ref{sec:BBwind}. To build on these results, we next (section \ref{sec:TeffkaprhoT}) extend the BB wind model to account for the temperature and density variation of the opacity, using the low-temperature opacity tables of \citet[][hereafter AF94, see figure \ref{fig:opal}]{Alexander94}; a key result, summarized in figure \ref{fig:TvsMdbv}, is that, because of the sharp drop in opacities due to electron recombination for $T \lesssim 6500$\,K, models with strong mass loss tend to converge to temperatures in the $5000-6000$\,K range inferred from the light echoes.
In section \ref{sec:D87anal} we revisit the D87 analysis, and show (figure \ref{fig:D87fig1}) that, for extension to the higher mass loss now inferred for $\eta$~Carinae, this model can also give spectral temperatures in the range inferred from the light echoes.
We conclude (section \ref{sec:concl}) with a brief comparative discussion of the merits of steady vs.\ explosive paradigms for modeling  giant eruptions in $\eta$\,Carinae and other LBVs, and provide an outlook for future work, including the broader application to highly optical thick outflows from LBVs in general, SN Type IIn precursors, and possibly classical nova eruptions.
%

\section{Black-Body Wind with Fixed Opacity}
\label{sec:BBwind}

The effective temperature $\Teff$ of a star is defined in terms of the surface flux, $ F= \sigma \Teff^4$, 
 where is $\sigma$ is the Stefan-Boltzmann constant.
For a stellar luminosity $L$ and surface radius $\Rstar$, such that $F=L/4 \pi \Rstar^2$, this can be scaled in terms of solar values,
\beq
\frac{L}{\Lsun} = 
\left ( \frac{\Rstar}{\Rsun} \right )^2 \,  \left ( \frac{\Teff}{\Tsun} \right )^4
\, .
\label{eq:lblsunbb} 
\eeq
Within a simple stellar atmosphere model with a gray (frequency-independent) opacity $\kappa$, the surface radius is defined as the location with radial optical depth $\tau(\Rstar)=2/3$, where
\beq
\tau (r) \equiv \int_r^\infty \kappa \rho(r') \, dr'
\, ,
\label{eq:taurdef}
\eeq
with $\rho$  the local mass density.
In this way the stellar effective temperature, which is formally defined as measure of the bolometric surface flux, is associated with an actual surface temperature $T_{\rm eff}= T(\tau = 2/3) $, which sets the overall spectral energy distribution and so represents a ``spectral temperature".

This gray-opacity association is normally applied for {\em hydrostatic} atmospheres, for which the density and pressure have a roughly exponential stratification with scale height 
$H \approx a^2/g$, where $g$ is the stellar surface gravity and $a$ is the isothermal sound speed,  with 
$H  \ll \Rstar$, implying a locally nearly planar atmosphere.

However for stars with very strong stellar wind outflows, the winds can become {\em optically thick}, implying that  the effective photosphere is now within the outflowing wind, with the radius of the ``wind-photosphere'' well above the hydrostatic stellar core radius, $\Rw \gg \Rstar$.
For a spherically symmetric, steady-state wind with local speed $V$, the total mass loss rate through any radius $r$ is $\Mdot \equiv 4 \pi  r^2 \rho V $. 
Solving for the local density $\rho$ and applying equation \ref{eq:taurdef}, we find that for a fixed opacity $\kappa$ and constant speed $V$ the integral can be trivially evaluated, giving an explicit expression for the associated radius where $\tau(\Rw)=2/3$, 
\beq
\Rw = \frac{3}{2} \, \frac{\kappa \Mdot}{4 \pi V} 
= 3665 \Rsun \, \frac{\kappa}{\kappa_e} \, \frac{\Mdot_{-2}}{V_8}
\, .
\label{eq:Rw}
\eeq
The latter equality here provides a numerical evaluation in terms of electron scattering opacity 
$\kappa_e = 0.34$\,cm$^2$\,g$^{-1}$, evaluated for fully ionized mixture with solar value for Hydrogen mass fraction $X=0.72$.
The wind parameters 
$\Mdot_{-2} \equiv (\Mdot/10^{-2} \Msun$/yr),
and 
$V_8 \equiv V/(10^8$\,cm/s) $=V/(1000$\,km/s)
are scaled by values characteristic of the extreme mass loss inferred for the giant eruptions of  LBV stars.

Applying this in equation (\ref{eq:lblsunbb}) and solving for the temperature gives\footnote{
This is essentially identical to equation (4) of \citet{Bath76}, used to interpret spectra of classical novae.},
\beqa
\Teff &=& \Tsun \, 
\left ( \frac{L}{\Lsun} \right )^{1/4} 
\left (  \frac{2}{3} \, \frac{4 \pi V \Rsun}{\kappa \Mdot}
\right )^{1/2}
\label{eq:Tke1}
\\
&=&
3030 \,{\rm K} ~ L_6^{1/4} 
\left (  \frac{\kappa_e}{\kappa} \frac{V_8} {\Mdot_{-2}}
\right )^{1/2}
\, ,
\label{eq:Tke}
\eeqa
where $L_6 \equiv L/10^6 \Lsun$.
The upshot here is that for sufficiently high mass loss rates, characteristic of what's inferred for giant eruption phases of LBVs and SN Type IIn precursors, the effective temperatures of even very luminous stars can become quite cool, in principle even cooler than the sun.\footnote{
On the other hand, classical nova eruptions at their peak are more ``intermediate". Their photosphere is already well above the acceleration zone, but because of the relatively lower mass loss rate, their typical photospheric temperature of 7000K is well above the ``opacity cliff" discussed below, with the coolest being about 6000K \citep{Hack93}.
}

However, an important caveat here is that, at such low temperatures, the recombination of Hydrogen and Helium greatly reduces the number of free electrons, making the effective opacity much lower than the value $\kappa_e=0.34$\,cm$^2$/g quoted above for electron scattering in a fully ionized plasma.
In \S \ref{sec:TeffkaprhoT} below we analyze how such an opacity reduction affects the scaling for effective temperature in this BB wind model.
But first the next section explores the effect of {\em sphericity} of the wind medium on the emergent spectral energy distribution.

\section{Spherical Gray Atmosphere in Radiative Equilibrium}
\label{sec:sgre}

To examine how the above idealized BB wind model for $T_{\rm eff}$ may be altered by sphericity effects,
let us consider now a spherical extension of the standard gray atmosphere in radiative equilibrium, which follows the LTE-like condition $S=J=B$,  with the spectrally integrated source function $S$ and mean intensity $J$ both set by the integrated Planck function $B$.
As reviewed in Chapter 19 of \citet[][hereafter HM14]{Hubeny14}, if we make the further assumption that the photon mean-free-path\footnote{Sometimes $n$ is characterized as an ``opacity" index, but since opacity can variously mean $\kappa \rho$ or $\kappa$, we choose the unambiguous characterization as a power index in the mean-free-path, $1/\kappa \rho$.} varies as a power-law in radius, $\ell = 1/\kappa \rho \sim r^n$, then the temperature variation in optical depth is given by HM14  equation (19.18),
\beq
T(\tau) = T_1 \tau^{1/2(n-1)} 
\left [ \frac{\tau + K_n}{1 + K_n} \right ]^{1/4}
\, ,
\label{eq:Ttau}
\eeq
where 
$K_n=(n+1)/3(n-1)$ is an integration 
constant\footnote{The value quoted comes from HM14 equation (19.18).
Note, however, that if the unit integration constant in HM14 (19.14) were replaced with 2, as chosen  in HM14 (19.12),
then $K_n$ would be a factor two higher. 
In the limit $n \rightarrow \infty$, this would give $K_n \rightarrow 2/3$, consistent with 
the usual (Eddington) value for a planar gray atmosphere.}.
Here also, $T_1 \equiv T(\tau=1)$, with radial optical depth
\beq
\tau (r) \equiv \int_r^\infty \kappa \rho \, dr' = \frac{C_n}{r^{n-1}(n-1)}
\, ,
\label{eq:taur}
\eeq
where $C_n$ is a normalization constant.
As also discussed in HM14, for a {\em non-radial} ray with impact parameter $p$ to the origin at $r=0$, the optical depth along a coordinate $z$ toward the observer (at $z \rightarrow \infty$) is given by
\beq
\tau(p,z) = \frac{C_n}{p^{n-1}} \int_0^{\arccos (z/r)} \sin^{n-2} (\theta) \, d\theta
\, ,
\label{eq:taupz}
\eeq
where $r=\sqrt{p^2+z^2}$; this integral can be evaluated analytically in terms of hypergeometric functions.
\begin{figure}
\begin{center}
\includegraphics[scale=0.5]{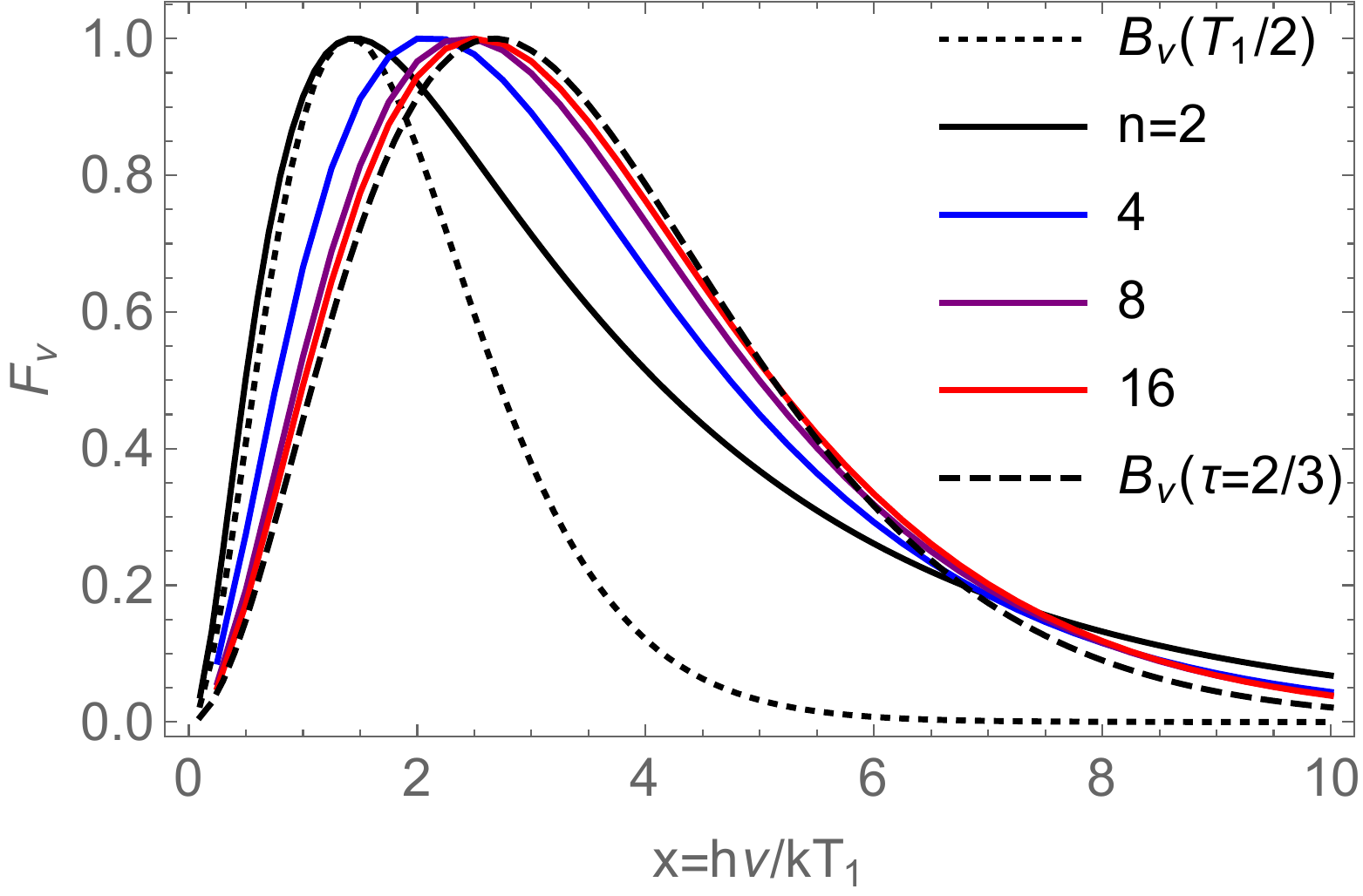}
\caption{Flux spectra for spherical, gray, radiative-equilibrium models with mean-free-path radial power-indices $n=$ 2, 4, 8 and 16, compared with Planck functions for the temperature at $\tau=2/3$ (dashed curve), and for $T=T_1/2$ (dotted curve).
To facilitate comparison of their relative forms, all curves have been normalized to have a peak value of unity.
}
\label{fig:fx}
\end{center}
\end{figure}

\begin{figure}
\begin{center}
\includegraphics[scale=0.6]{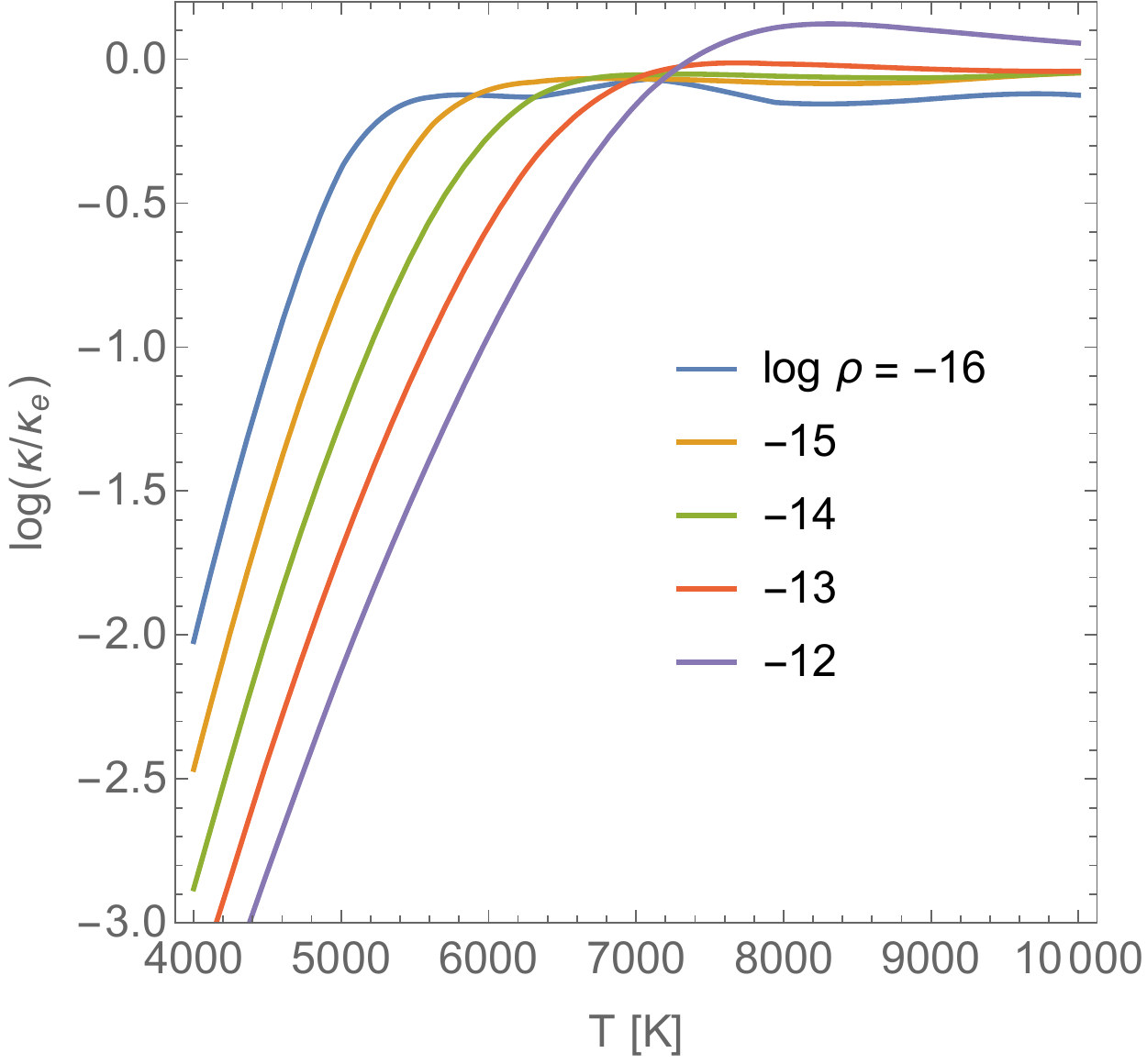}
\caption{
For the AF94 opacity tables,
plots of $\log(\kappa/\kappa_e )$ vs.\ temperature $T$ for selected densities over the logarithmic range $\log \rho =$ -16 to -12 (g/cm$^{3}$).
The sharp drop for $T<6500$\,K shows the ``opacity cliff", which has a major effect in clustering wind effective temperatures around $\Teff \approx 5500$\,K for a wide range of luminosity and mass-loss-rate (see figure \ref{fig:TvsMdbv}).
}
\label{fig:opal}
\includegraphics[scale=0.5]{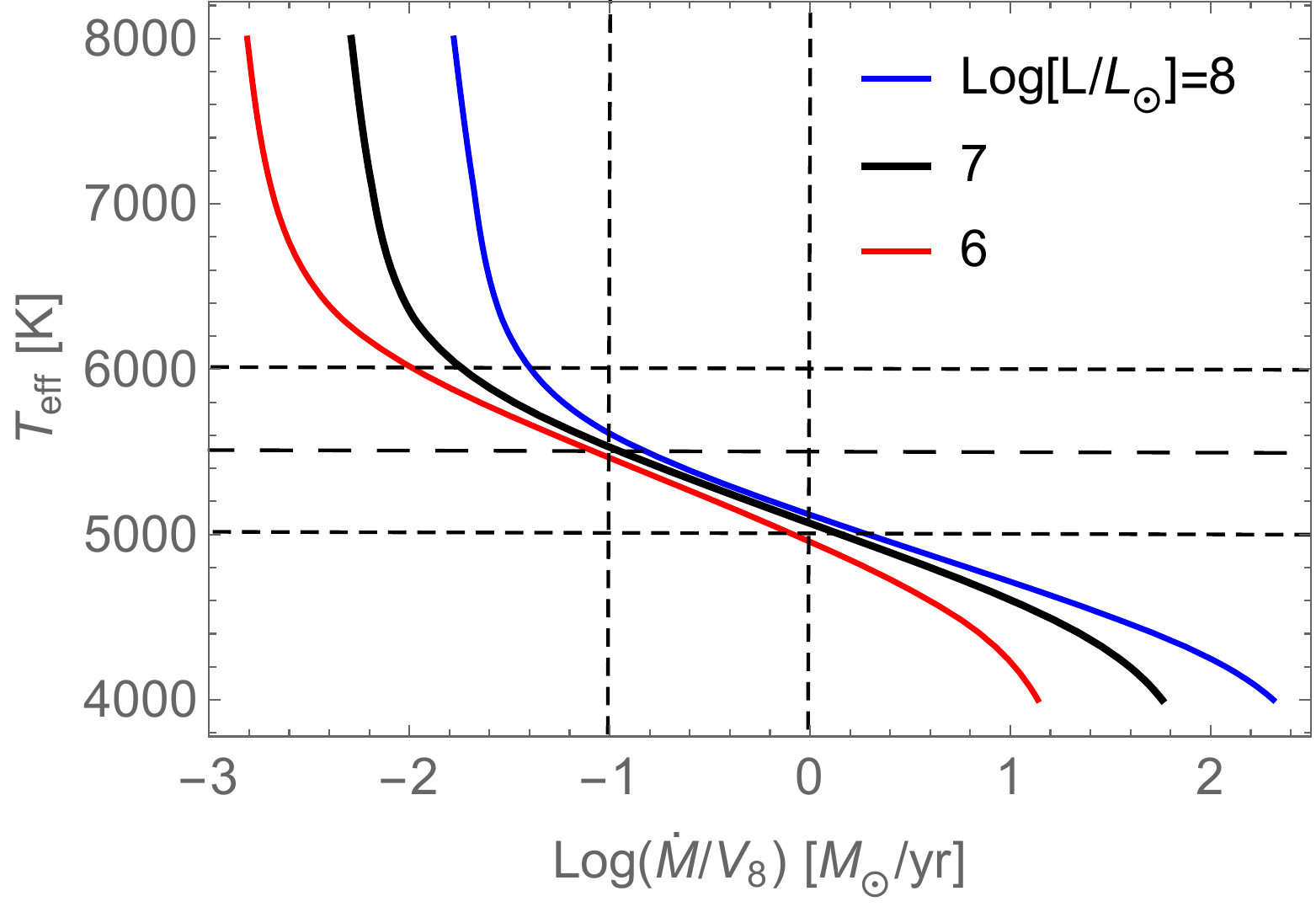}
\caption{
For the AF94 opacities applied to an LTE blackbody model of an optically thick wind, self-consistent solution for effective temperature $\Teff$ (in K) plotted vs.\ mass-loss-to-velocity ratio $\Mdot/V_8$ (on a log scale), for selected luminosities  $\log (L/\Lsun) =$\, 6, 7 and 8.
The dashed lines highlight the narrow range of solutions with $\Teff = 5000-5500$\,K over a 1 dex range in $\Mdot/V_8$.
This clustering of $\Teff$ around 5500\,K is a direct consequence of the sharp ``opacity cliff" for temperatures $T<6500$\,K, as shown in figure \ref{fig:opal}, and discussed in the text.}
\label{fig:TvsMdbv}
\end{center}
\end{figure}

The associated observed intensity $I_\nu (p,\infty)$ can then be obtained by numerical evaluation of the formal solution integral given in HM14 equation (19.20). With further numerical evaluation of the integral (HM14 equation 19.21) over impact parameter $p$, we obtain the emergent flux  spectrum $F_\nu$ vs. frequency $\nu$, which can be conveniently cast in terms of the scaled frequency $x \equiv h \nu/kT_1$.

Figure \ref{fig:fx} plots the emergent flux spectra (normalized to have peak flux of unity) for various mfp power-indices $n =$ 2, 4, 8 and 16. The dashed curve is the usual emergent spectrum from a planar gray atmosphere, closely set by the Planck function at the optical depth $\tau=2/3$.
Note that the lower $n$ models have a peak that is shifted to lower frequency, representing then a lower spectral temperature.
In particular, the peak for $n=2$ can be fit roughly with that of a Planck function with temperature $T \approx 0.5 T_1$ (cf.\ the solid black and dotted curves in figure \ref{fig:fx}), but the high energy tail is distinctly stronger than for a Planck function of this reduced temperature.
With increasing $n$, sphericity effects become ever less pronounced, and so at large $n$ we roughly recover the planar atmosphere result set by $B_\nu (\tau=2/3)$, as given in the dashed curve.

The overall implication is that, within such a gray, radiative equilibrium model, sphericity effects lead to a spectral temperature that is generally moderately lower than inferred from the simple BB wind model, but which fully recovers the BB result in the planar atmosphere limit $n \rightarrow \infty$.

\section{Effective temperature for blackbody wind with $\kappa(\rho,T)$}
\label{sec:TeffkaprhoT}

Let us next extend the simple BB wind model of section \ref{sec:BBwind} 
to account for dependence of the opacity on temperature and density,
$\kappa (\rho, T)$, as given here from the low-temperature Rosseland-mean opacity tables compiled by AF94.
Figure \ref{fig:opal} shows a line plot of $\log[\kappa(\rho,T)/\kappa_e]$ vs. $T$ for selected values of density  $\log \rho$.
For moderately high temperatures above about 6500\,K, the opacity is close to the electron scattering value, $\kappa_e = 0.34$\,cm$^2$/g;
but for lower temperatures, note how it drops sharply, so that for $T=4000$\,K it is reduced by 2-3 dex compared to the electron scattering value.
This low-T ``{\em opacity cliff}" has important implications for the effective temperature of very dense wind outflows.

To solve self-consistently for the effective BB wind temperature accounting for this temperature- and density-dependent opacity, we follow a straightforward iterative approach.
First, given values of luminosity and effective temperature, equation (\ref{eq:lblsunbb}) can be solved for an effective ``surface'' radius in the wind, $\Rstar \rightarrow \Rw$, which, for given ratio of mass loss rate to wind speed, $\Mdot/V$, yields an associated surface density $\rho_w = \Mdot/4\pi V R_w^2$.
Then applying this and the assumed temperature to evaluate the opacity in equation (\ref{eq:Rw}), we solve iteratively for $\Teff$ for given values of luminosity $L$ and the mass-loss-rate to speed ratio $\Mdot/V$.
The convergence is typically quite fast, requiring only a few ($<$\,10) iterations to reach a self-consistent model.

Figure \ref{fig:TvsMdbv} plots the resulting variation of $\Teff$ vs.\ $\Mdot/V_8$ 
(in $\Msun$/yr),
for  luminosities $\log(L/\Lsun ) =$\, 6, 7, and 8.
For low values of the ratio $\Mdot/V_8$, we find high temperatures, $\Teff > 6000$\,K, for which the $\kappa \approx \kappa_e$,  so that the solutions closely follow analytic scaling form of equation (\ref{eq:Tke}).

But for increasing  $\Mdot/V_8$, for which the effective temperature drops {\em below} 6000\,K, the sharp reduction in opacity means that the effective radius becomes nearly fixed, leading then to an effective temperature in the narrow range $\Teff  = 5500 \pm 500 $\,K for  all three luminosities and over more than an order magnitude range in mass-loss-to-speed ratio $\Mdot/V_8$.

\begin{figure}
\includegraphics[scale=0.42]{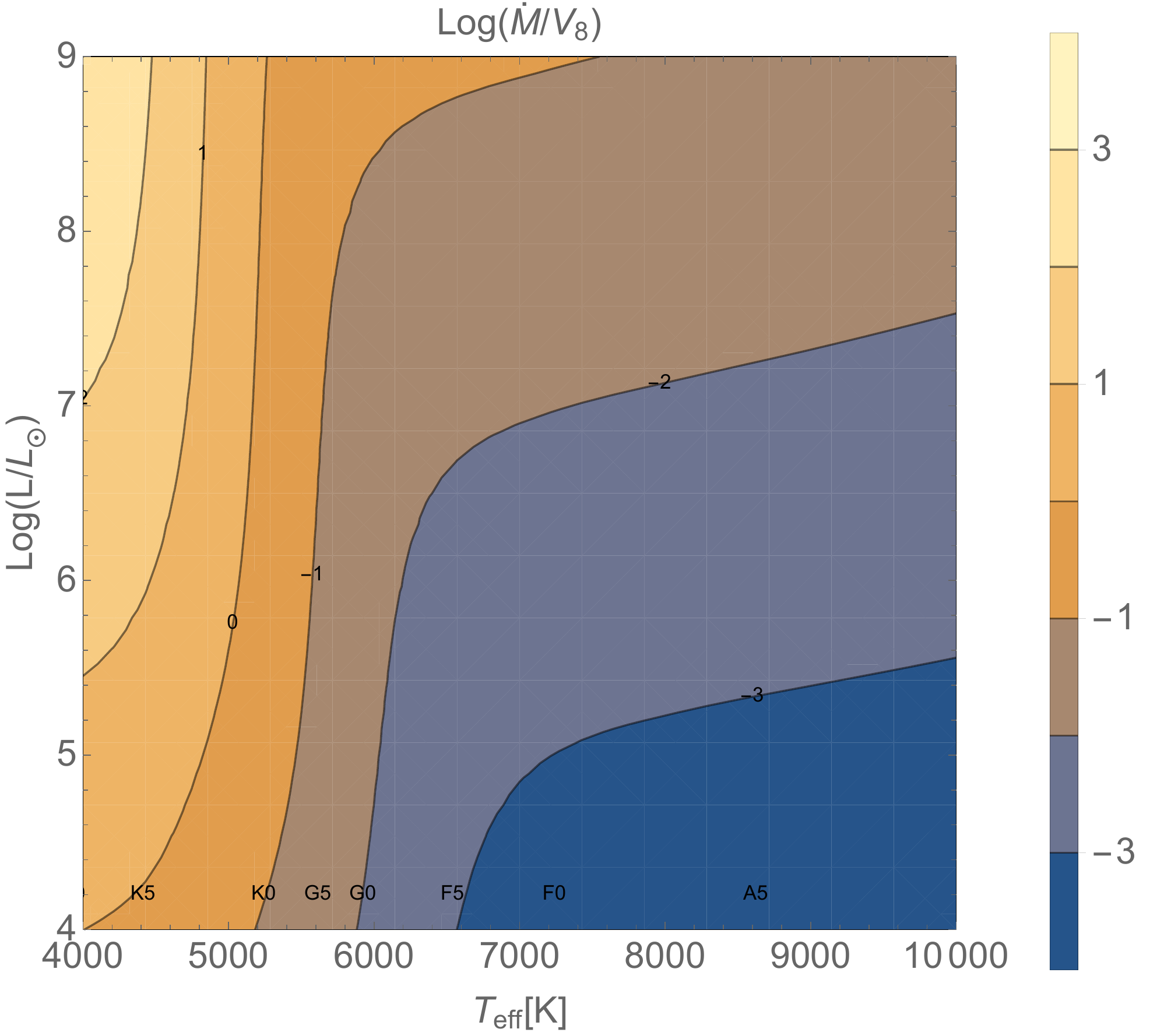}
\caption{
Contours of $\log \Mdot/V_8$ vs. $\log(L/\Lsun)$ and $T_{\rm eff}$ for the BB-wind model with AF94 opacities.
The steep vertical contours for $T_{\rm eff}=5000-6000$\,K  show the effect of the opacity cliff in this temperature range.  Thus, the temperature is not a good diagnostic of mass loss around the opacity cliff, but it is useful at higher temperatures, $T_{\rm eff} \gtrsim 6500$\,K.
}
\label{fig:MdbvLT}
\end{figure}

For this BB wind model with AF94 opacities, figure \ref{fig:MdbvLT} plots contours of $\log (\Mdot/V_8)$ vs.\ $\log (L/\Lsun )$ and $T_{\rm eff}$. The steep vertical contours for $T_{\rm eff}=5000-6000$\,K again show the effect of the opacity cliff in this temperature range.

The upshot here is that, because of the abrupt opacity cliff associated with the recombination of electrons at low temperature, the effective temperature of this BB wind tends to become fixed in the 5000-6000\,K range, thus corresponding quite naturally to the temperature inferred from light echo spectra of the giant eruption of $\eta$\,Carinae.



\begin{figure*}
\begin{center}
\includegraphics[scale=0.65]{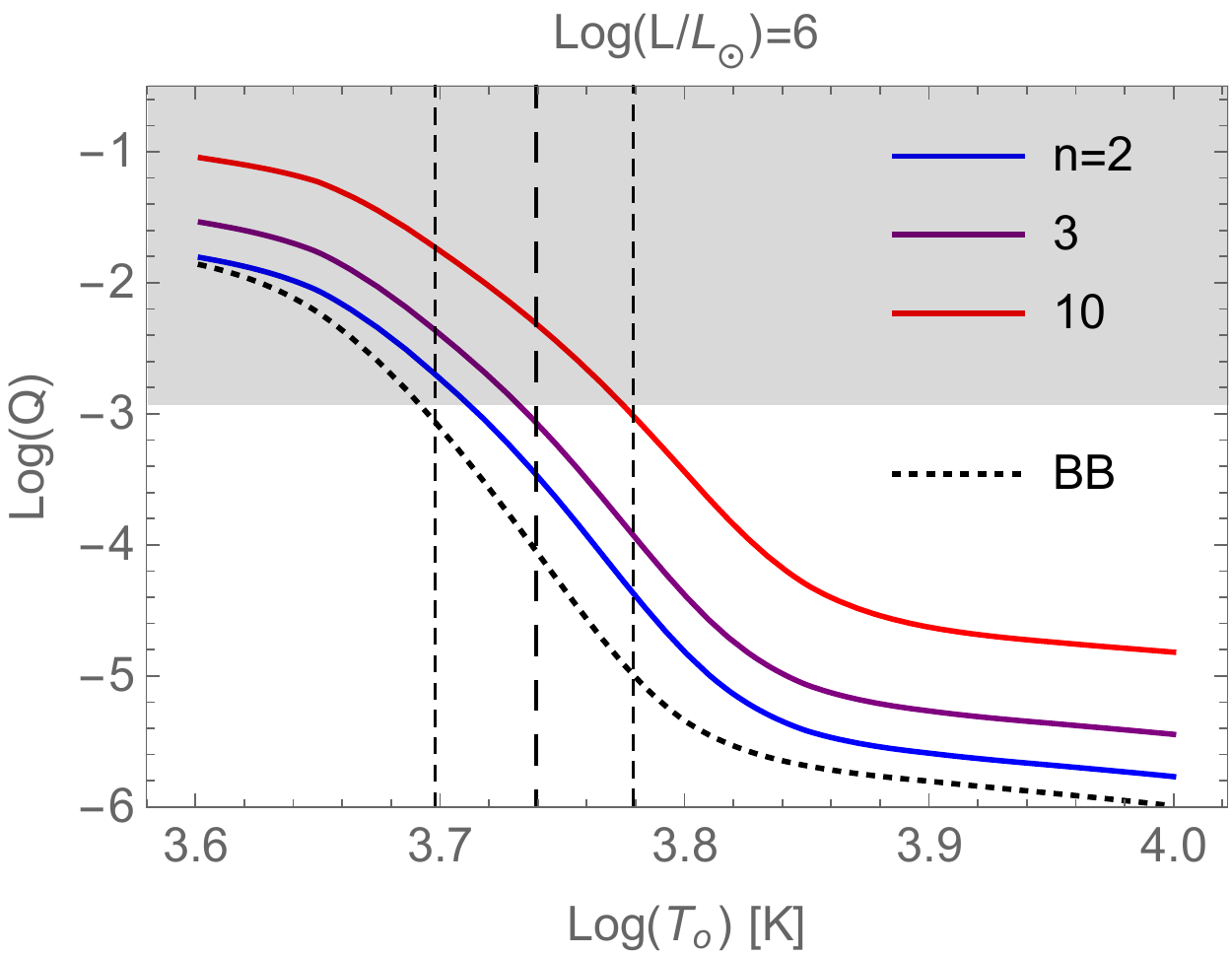}
~~~~~~~~~~
\includegraphics[scale=0.65]{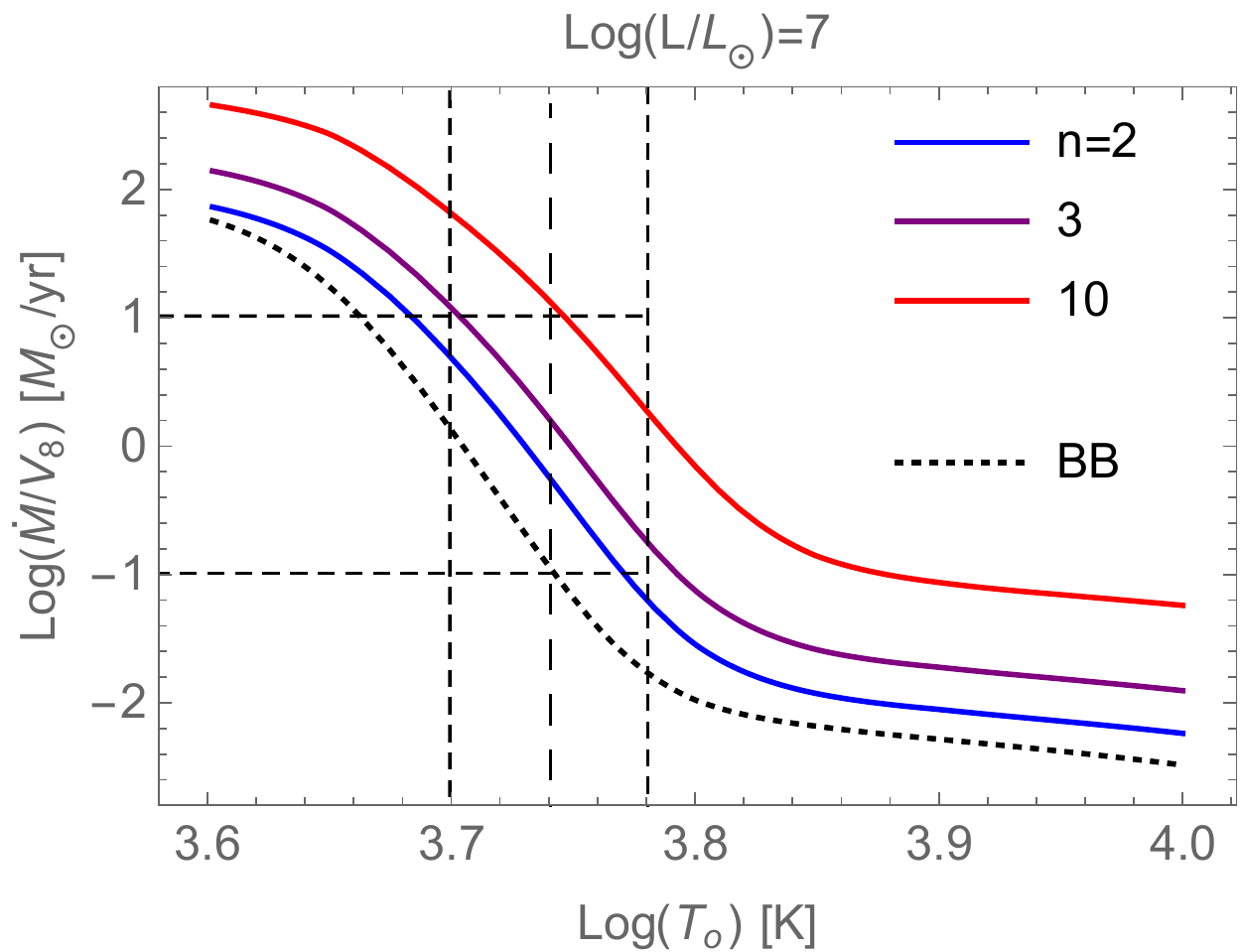}
\caption{
{\em Left:}
For an assumed luminosity $L=10^6 \Lsun$, our reproduction of D87 figure 1, using now the AF94 opacities, and extending to much larger Q values (gray area), reflecting the much higher mass loss rates now inferred for the giant eruption of $\eta$\,Carinae. 
The  dotted curve represent corresponding results for the BB-wind model.
To focus on the low-temperature range 5000-6000\,K (marked by the vertical dashed lines), we limited the maximum temperature here to $\log(T_o)=4$.
For the $\log(Q)<-3$ range (below the gray) assumed in D87, the agreement between these and D87 results is quite good, indicating that the opacities assumed by D87 are quite similar to the AF94 opacities taken here.
{\em Right:}
For the same set of models shown in the left panel, but with a higher luminosity $L=10^7 \Lsun$ characteristic an LBV giant eruption, 
line plots of mass-loss-by-speed parameter $\Mdot/V_8$ vs.\ temperature $T_o$.
The horizontal dashed lines show that achieving the median cool temperature of 5500\,K requires a relatively moderate $\Mdot/V_8 = 0.1 \Msun/$yr in the BB wind model (dotted curve), but more than a factor {\em hundred} higher mass loss in the D87 model with $n=10$ (red curve). }
\label{fig:D87fig1}
\end{center}
\end{figure*}

\section{Extending the D87 Analysis}
\label{sec:D87anal}

\subsection{Method and original results}
\label{sec:D87orig}

In one of the first attempts to estimate the spectral temperature of LBV eruptions, \citet[][D87]{Davidson87} used escape probability arguments to derive scaling equations for the temperature $T_o$ at a surface radius $r_o$.
To account for sphericity, this again  assumes the photon mean-free-path scales as  a simple power-law in radius, viz., $\ell \equiv 1/\kappa \rho \sim r^n$.
The analysis is likewise based on a mean, effectively gray opacity; but instead of assuming radiative equilibrium, it invokes a true-absorption opacity -- defined as a fraction $\alpha_o$ of the total  opacity $\kappa_o$ -- that {\em reduces} the local luminosity (see D87 equation A2).
As we discuss in section \ref{sec:D87vsBB}, this represents a key difference from the gray, radiative equilibrium model given in section \ref{sec:sgre} above.

Through an effective optical depth integral, equation (2) of D87 relates the opacity $\kappa_o$ and density $\rho_o$ at the reference radius $r_o$,
\beq
\alpha_o^{1/2} \kappa_o \rho_o r_o = A_n
\, ,
\label{eq:D87-2}
\eeq
where the coefficient $A_n$ depends on the power index $n$.
D87 equation (3) then relates the luminosity to the temperature $T_o$ at this radius $r_o$,
\beq
L = 4 B_n \, 4 \pi r_o^2 \sigma T_o^4 
\, ,
\label{eq:D87-3}
\eeq
where $B_n$ is another coefficient that depends on $n$, and the Stefan-Boltzmann constant $\sigma$ is related to the radiation constant $a$ and speed of light $c$ through $\sigma = ac/4$.

Note that setting $\alpha_o = 1$, $A_n=2/3$, and $B_n=1/4$ {\em recovers the BB-wind equations} of section \ref{sec:BBwind}.

For general $n$ and associated $A_n$ and $B_n$, 
D87 claim that solutions for temperature $T_o$ depend mainly on a single parameter (see their equation 4) defined\footnote{This 
differs somewhat from the $T \sim L^{-1/2}$ scaling predicted in equation (\ref{eq:Tke1} ) from the BB wind model of section \ref{sec:BBwind}.
In practice we find here that there is some residual extra dependence on $L$ for the parameter $Q$, so in figure \ref{fig:D87fig1} we assume a value $\log(L/\Lsun )=6$ that seems to best match D87 figure 1.}
by $Q \equiv \Mdot/v_o/L_6^{0.7}$, with $v_o$ in km/s and $\Mdot$ in $\Msun$/yr.
For fixed $\alpha_o = 0.3$, D87 figure 1 plots solutions for fixed $n=$\,2, 3, and 10 in the $\log(Q)$ vs.\ $T_o$ plane. 
A key result is that as temperature declines to around 7000\,K, the drop in opacity leads to a sharp upturn to large $Q$.

D87 thus stated that, ``even if $\Mdot$ becomes very large, $T_o$ cannot easily fall below 7000\,K". 
This was cited by \citet{Rest12} in arguing that the lower 5000-6000\,K temperatures inferred from light echoes of $\eta$\,Carinae suggests an incompatibility with a steady wind model, and thus may instead point to a more explosive scenario for such giant eruptions.

\subsection{Extensions to higher mass-loss rate}
\label{sec:D87ext}

Our extensions here of this D87 analysis explore cases up to $\log Q =-1$, two dex higher than in D87;
this is to include the higher eruptive mass loss rate implied by the large ($>10 \Msun$) mass now inferred for the Homunculus nebula
 \citep{Smith03b}.
We retain the assumption of fixed $\alpha_o =0.3$, 
using now the AF94 low-$T$ tabulations for the opacity $\kappa$.
%
For an assumed luminosity\footnote{This was chosen to most closely match the fixed-$n$ curves of D87, which doesn't specifically state the luminosity used to make their figure 1.} $\log(L/\Lsun) = 6$, the left panel of figure \ref{fig:D87fig1} reproduces and extends results from D87 figure 1,
plotting $\log Q$ vs. $T_o$ for the same fixed index models $n=$\,2, 4, and 10 (blue, purple, red), as well for the BB-wind (dotted curve).

For a higher luminosity $\log(L/\Lsun ) = 7$ that is more appropriate for giant eruption LBV's, the right panel of figure  \ref{fig:D87fig1} plots directly the mass-loss to speed ratio $\Mdot/V_8$ vs. $T_o$ for the same set of models shown in the left panel. A key result is that all models can give solutions in the $5000-6000$\,K range (marked by vertical dashed lines) inferred from $\eta$\,Carinae's light echoes \citep{Rest12}, but the D87 models require much higher mass loss rates (by up to a factor 100 for the $n=10$ model) than the simple BB wind model (dotted curve) derived here.

Despite this wide range, an important consequence here is that even in the D87 models, the temperature can readily drop below the previously cited limit of 7000\,K, and for mass loss rates in the observationally inferred range, even approach the low $T \approx 5500$\,K spectral temperature inferred by \citet{Rest12} from light echo spectra of $\eta$\,Carinae.
This means that such wind expansion models are not inherently problematic for explaining $\eta$\,Carinae's giant eruption, and the lower spectral temperature inferred from its light echoes.
Indeed, within the context of the BB wind model with the AF94 opacities, one should actually have expected a temperature in this range for a quite wide range of mass loss conditions.

\subsection{Contrast with spherical gray models}
\label{sec:D87vsBB}
Let us next directly contrast these D87 results with those obtained from the spherical, gray, radiative equilibrium analysis in section \ref{sec:sgre}.
As noted there, for a gray atmosphere requiring radiative equilibrium implies an LTE-like condition $S=J=B$, regardless of whether the opacity is of a scattering or pure-absorption type.  This would thus correspond to taking $\alpha_o=1$ in the D87 approach, but that would represent only a modest difference from the fixed $\alpha_o = 0.3$ assumed in all the models presented in D87.

Much more significant are the distinct differences in the derivation of an effective spectral temperature from a spherical model with assumed radial power index $n$ for the mean-free-path. In the gray, radiative equilibrium analysis, this leads to a specific relation (equation \ref{eq:Ttau}) for the optical depth variation of temperature, which is then used to derive formal solutions for the emergent spectrum. Two key results are: (1) sphericity tends to {\em lower} the frequency of the spectral peak, implying a {\em reduced} spectral temperature; (2)  however, in the limit of large $n$, one recovers the standard blackbody result $B_\nu(\tau=2/3)$ of a planar gray atmosphere. Since this was the basis of the simple BB wind model in section \ref{sec:BBwind}, it suggests that the extension of the BB wind model to account for realistic opacity variations should give a pretty good approximation of the spectral temperatures of such very optically thick winds, with the results in figure \ref{fig:TvsMdbv} providing a good prediction for their variation with mass-loss rate.

In contrast, in the D87 approach there is no attempt to model the optical depth variation of the temperature. Instead, one uses escape probability arguments to estimate the characteristic energy of the photons that escape through the effective ``surface'' radius $r_o$, under the different indices $n$ that represent the steepness of the variation in mean-free-path. A key prediction now is that, for increasing steepness represented by higher $n$, the emergent temperature becomes {\em higher}, apparently because higher energy photons (associated with the higher intensity of the less-attenuated inner luminosity) can then more easily escape through the much narrower surface layers.
This stands in direct opposition to the {\em lower} spectral temperature predicted from the spherical, radiative equilibrium model, which can be viewed as stemming from a marked enhancement in the center-to-limb gravity darkening for a spherical wind with lower index $n$.

To decide between these distinct predictions, we note here that there seems an inherent inconsistency in D87's attempt to account for scattering vs.\ pure-absorption effects within an analysis that does not account for the frequency variation of opacity. In particular, their equation (A2) directly contradicts the normal requirement of radiative equilibrium, viz.\ that the bolometric luminosity must remain constant.  The assumption instead that this luminosity is reduced by the absorption component of opacity overlooks how and where that absorbed energy reappears, as it must in radiative equilibrium. This would seem to call into question the physical basis for the predictions of higher spectral temperatures with increasing $n$.

We thus argue here that, within the current state of analyses of such very optically thick outflows, the BB wind with variable opacity, and the spherical, gray, radiative equilibrium models, provide the best available predicted scalings for the spectral temperature and associated flux spectrum.
But ultimately the validity this approach, and the issues surrounding the D87 escape analysis, should be examined with more fundamental models that account for non-LTE effects with a non-gray opacity within a spherically expanding wind.

\section{Concluding Discussion}
\label{sec:concl}

A key specific result of the steady-state, optically thick wind outflow analyses here is that the low temperatures in the range $5000-6000$\,K inferred by \citet{Rest12} from light echoes of 
$\eta$\,Carinae's giant eruption are in fact compatible, and indeed predicted, for a quite wide range of very large mass-loss rates ($0.01 - 1 \, \Msun$/yr) and luminosities ($ L/\Lsun = 10^6 - 10^8 $; see figure \ref{fig:TvsMdbv}) that are consistent with values estimated for the 1840's eruption epoch.


But it is important to emphasize here that such a cool spectral temperature is {\em not unique} to such a steady wind outflow model. In particular, Type IIP supernovae (SN) {\em explosions} show extended periods with a spectral temperature in this 5000-6000\,K range, attributed to formation of a ``recombination photosphere" \citep{Shigeyama91, Dessart11}, again associated with the loss of free electron scattering opacity due to hydrogen recombination in this temperature range. In effect, the steep opacity cliff shown in figure \ref{fig:TvsMdbv} regulates the effective spectral temperature in both steady and explosive models to remain at the recombination temperature that characterizes this opacity drop. As such, inferring such a low spectral temperature from an LBV giant eruption does not, by itself, help much to discriminate between an explosion vs\ steady-wind scenario.

Indeed, in addition to  the low spectral temperature, \citet{Rest12} and a follow up study of the light echoes by  \citet{Prieto14}, 
 provide several additional arguments  against a steady wind and in favor of an explosive origin of $\eta$\,Carinae's giant eruption.
Similar and additional arguments \citep{Smith10,Smith11,Smith14} 
 favoring an explosive scenario have also been given for the broader population of giant-eruption LBVs.

In the context of the applicability of the steady wind outflow models here, a key question is whether the timescale for variations in the outflow is longer than a characteristic flow time, which for the flow speeds and extended wind-photospheric radii of eruptive LBV's can range up to several months.
For cases like $\eta$\,Carinae that show an extended epoch of outburst extending or years and even a decade, a steady model can represent a useful idealization. The opposite limit of a sudden explosion, as classically modeled from similarity, Sedov-type solution,  provides a counter idealization. But a more realistic model may draw from insights gleaned from both ideal limits. Similar arguments apply for pre-supernovae eruptions, and for novae.

Within the steady-wind approach, there is a strong need now to carry out more complete models that account for non-LTE effects with a non-gray opacity within a spherically expanding wind.
By deriving detailed spectra, this would test the importance of non-LTE effects and deviations from a gray LTE model, and resolve definitively the differences between the D87 and BB wind scalings for spectral temperature.
We hope next to carry out such modeling using one of the non-LTE wind codes mentioned in the introduction, with focus on extending their opacity treatments to the lower temperature conditions expected from strong giant eruption phases of LBV stars.


\section*{Acknowledgments}

SPO acknowledges partial support by a U.S. NSF grant \#1312898, awarded to the University of Delaware.
NJS acknowledges partial support by an Israeli Pazi grant by the IAEC and PBC and by the I-CORE Program of the PBC and ISF (center 1829/12).
We thank Ivan Hubeny for helpful discussions on the spherical gray model and Chapter 19 of his book.
We also thank Nathan Smith and John Hillier for constructive criticisms and suggestions from reading an early draft.
  
\bibliographystyle{mn2e}
\bibliography{OwockiS}

\begin{thebibliography}{}

\bibitem[\protect\citeauthoryear{{Alexander} \& {Ferguson}}{{Alexander} \&
  {Ferguson}}{1994}]{Alexander94}
{Alexander} D.~R.,  {Ferguson} J.~W.,  1994, \apj, 437, 879

\bibitem[\protect\citeauthoryear{{Bath} \& {Shaviv}}{{Bath} \&
  {Shaviv}}{1976}]{Bath76}
{Bath} G.~T.,  {Shaviv} G.,  1976, \mnras, 175, 305

\bibitem[\protect\citeauthoryear{{Davidson}}{{Davidson}}{1987}]{Davidson87}
{Davidson} K.,  1987, \apj, 317, 760

\bibitem[\protect\citeauthoryear{{de Koter}, {Lamers} \& {Schmutz}}{{de Koter}
  et~al.}{1996}]{deKoter96}
{de Koter} A.,  {Lamers} H.~J.~G.~L.~M.,    {Schmutz} W.,  1996, \aap, 306, 501

\bibitem[\protect\citeauthoryear{{Dessart} \& {Hillier}}{{Dessart} \&
  {Hillier}}{2011}]{Dessart11}
{Dessart} L.,  {Hillier} D.~J.,  2011, \mnras, 410, 1739

\bibitem[\protect\citeauthoryear{{Friedjung}}{{Friedjung}}{2011}]{Friedjung11}
{Friedjung} M.,  2011, \aap, 536, A97

\bibitem[\protect\citeauthoryear{{Hack}, {Selvelli} \& {Duerbeck}}{{Hack}
  et~al.}{1993}]{Hack93}
{Hack} M.,  {Selvelli} P.,    {Duerbeck} H.~W.,  1993, NASA Special
  Publication, 507

\bibitem[\protect\citeauthoryear{{Hillier}}{{Hillier}}{2012}]{Hillier12}
{Hillier} D.~J.,  2012, in {Richards} M.~T.,  {Hubeny} I.,  eds, From
  Interacting Binaries to Exoplanets: Essential Modeling Tools Vol.~282 of IAU
  Symposium, {Hot Stars with Winds: The CMFGEN Code}.
pp 229--234

\bibitem[\protect\citeauthoryear{{Hillier} \& {Miller}}{{Hillier} \&
  {Miller}}{1999}]{Hillier99}
{Hillier} D.~J.,  {Miller} D.~L.,  1999, \apj, 519, 354

\bibitem[\protect\citeauthoryear{{Hubeny} \& {Mihalas}}{{Hubeny} \&
  {Mihalas}}{2014}]{Hubeny14}
{Hubeny} I.,  {Mihalas} D.,  2014, {Theory of Stellar Atmospheres}

\bibitem[\protect\citeauthoryear{{Humphreys} \& {Davidson}}{{Humphreys} \&
  {Davidson}}{1994}]{Humphreys94}
{Humphreys} R.~M.,  {Davidson} K.,  1994, \pasp, 106, 1025

\bibitem[\protect\citeauthoryear{{Ofek}, {Sullivan}, {Shaviv}, {Steinbok},
  {Arcavi}, {Gal-Yam}, {Tal}, {Kulkarni}, {Nugent}, {Ben-Ami}, {Kasliwal},
  {Cenko}, {Laher}, {Surace}, {Bloom}, {Filippenko}, {Silverman} \&
  {Yaron}}{{Ofek} et~al.}{2014}]{Ofek14}
{Ofek} E.~O.,  {Sullivan} M.,  {Shaviv} N.~J.,  {Steinbok} A.,  {Arcavi} I.,
  {Gal-Yam} A.,  {Tal} D.,  {Kulkarni} S.~R.,  {Nugent} P.~E.,  {Ben-Ami} S.,
  {Kasliwal} M.~M.,  {Cenko} S.~B.,  {Laher} R.,  {Surace} J.,  {Bloom} J.~S.,
  {Filippenko} A.~V.,  {Silverman} J.~M.,    {Yaron} O.,  2014, \apj, 789, 104

\bibitem[\protect\citeauthoryear{{Pauldrach}, {Hoffmann} \&
  {Lennon}}{{Pauldrach} et~al.}{2001}]{Pauldrach01}
{Pauldrach} A.~W.~A.,  {Hoffmann} T.~L.,    {Lennon} M.,  2001, \aap, 375, 161

\bibitem[\protect\citeauthoryear{Prieto, Rest, Bianco, Matheson, Smith,
  Walborn, Hsiao, Chornock, {\'A}lvarez, Campillay et~al.,}{Prieto
  et~al.}{2014}]{Prieto14}
Prieto J.,  Rest A.,  Bianco F.,  Matheson T.,  Smith N.,  Walborn N.,  Hsiao
  E.,  Chornock R.,  {\'A}lvarez L.~P.,  Campillay A.,    et~al., 2014, The
  Astrophysical Journal Letters, 787, L8

\bibitem[\protect\citeauthoryear{{Puls}, {Urbaneja}, {Venero}, {Repolust},
  {Springmann}, {Jokuthy} \& {Mokiem}}{{Puls} et~al.}{2005}]{Puls05}
{Puls} J.,  {Urbaneja} M.~A.,  {Venero} R.,  {Repolust} T.,  {Springmann} U.,
  {Jokuthy} A.,    {Mokiem} M.~R.,  2005, \aap, 435, 669

\bibitem[\protect\citeauthoryear{{Puls}, {Vink} \& {Najarro}}{{Puls}
  et~al.}{2008}]{Puls08}
{Puls} J.,  {Vink} J.~S.,    {Najarro} F.,  2008, \aapr, 16, 209

\bibitem[\protect\citeauthoryear{{Rest}, {Prieto}, {Walborn}, {Smith},
  {Bianco}, {Chornock}, {Welch}, {Howell}, {Huber}, {Foley}, {Fong}, {Sinnott},
  {Bond}, {Smith}, {Toledo}, {Minniti} \& {Mandel}}{{Rest}
  et~al.}{2012}]{Rest12}
{Rest} A.,  {Prieto} J.~L.,  {Walborn} N.~R.,  {Smith} N.,  {Bianco} F.~B.,
  {Chornock} R.,  {Welch} D.~L.,  {Howell} D.~A.,  {Huber} M.~E.,  {Foley}
  R.~J.,  {Fong} W.,  {Sinnott} B.,  {Bond} H.~E.,  {Smith} R.~C.,  {Toledo}
  I.,  {Minniti} D.,    {Mandel} K.,  2012, \nat, 482, 375

\bibitem[\protect\citeauthoryear{{Shigeyama} \& {Nomoto}}{{Shigeyama} \&
  {Nomoto}}{1991}]{Shigeyama91}
{Shigeyama} T.,  {Nomoto} K.,  1991, in {Woosley} S.~E.,  ed., Supernovae
  {Lightcurve Models for Supernova 1987A - Roles of the Recombination Front of
  Hydrogen}.
p.~198

\bibitem[\protect\citeauthoryear{{Smith}}{{Smith}}{2014}]{Smith14}
{Smith} N.,  2014, \araa, 52, 487

\bibitem[\protect\citeauthoryear{{Smith}, {Bally} \& {Walborn}}{{Smith}
  et~al.}{2010}]{Smith10}
{Smith} N.,  {Bally} J.,    {Walborn} N.~R.,  2010, \mnras, 405, 1153

\bibitem[\protect\citeauthoryear{{Smith}, {Gehrz}, {Hinz}, {Hoffmann}, {Hora},
  {Mamajek} \& {Meyer}}{{Smith} et~al.}{2003}]{Smith03b}
{Smith} N.,  {Gehrz} R.~D.,  {Hinz} P.~M.,  {Hoffmann} W.~F.,  {Hora} J.~L.,
  {Mamajek} E.~E.,    {Meyer} M.~R.,  2003, \aj, 125, 1458

\bibitem[\protect\citeauthoryear{{Smith}, {Li}, {Silverman}, {Ganeshalingam} \&
  {Filippenko}}{{Smith} et~al.}{2011}]{Smith11}
{Smith} N.,  {Li} W.,  {Silverman} J.~M.,  {Ganeshalingam} M.,    {Filippenko}
  A.~V.,  2011, \mnras, 415, 773

\end{thebibliography}

\end{document}